\documentclass{ws-procs975x65}
\usepackage{graphicx}

\def\beq{\begin{equation}}
\def\eeq{\end{equation}}

\begin{document}

\title{Adiabatic and post-adiabatic approaches to extreme mass ratio inspiral}
\author{Scott A.\ Hughes$^*$}

\address{Department of Physics and MIT Kavli Institute, MIT,
  Cambridge, MA 02139, USA\\
$^*$E-mail: sahughes@mit.edu}

\begin{abstract}
Extreme mass ratio inspirals (EMRIs) show a strong separation of
timescales, with the time characterizing inspiral, $T_{\rm i}$, much
longer than any time $T_{\rm o}$ characterizing orbital motions.  The
ratio of these timescales (which is essentially an EMRI's mass ratio)
can be regarded as a parameter that controls a perturbative expansion.
Here we describe the value and limitations of an ``adiabatic''
description of these binaries, which uses only the leading terms
arising from such a two-timescale expansion.  An adiabatic approach
breaks down when orbits evolve through resonances, with important
dynamical and observational consequences.  We describe the shortfalls
of an approach that only includes the adiabatic contributions to EMRI
evolution, and outline what must be done to evolve these systems
through resonance and to improve our ability to model EMRI systems
more generally.
\end{abstract}


\bodymatter


\section{Motivation: The large-mass ratio limit of the two-body problem
and extreme mass ratio inspirals}

Binary systems in which one body is much more massive than the other
can be analyzed perturbatively.  We can describe such a binary as an
exact black hole solution of general relativity (corresponding to the
larger member of the binary) plus a correction due to the smaller
body.  Because the perturbation equations are much simpler to solve
than the complete equations of general relativity, this turns out to
be a limit that can be modeled very accurately and precisely.

At least two major science goals drive studies of large mass ratio
systems.  First, these binaries represent a limit of the two-body
problem that can be solved with high precision.  As such, the study of
these binaries provides important input to programs to solve the
two-body problem of general relativity more generally, such as
numerical relativity and the effective one-body approach\cite{numrel1,
  numrel2, eob}.  Second, astrophysical extreme mass ratio inspirals
(EMRIs) are expected to be important sources for space-based GW
detectors such as eLISA\cite{eLISA} and DECIGO\cite{decigo}.

In this article, we will focus on the role of EMRIs as sources of
gravitational waves (GWs).  Such binaries are created when multibody
interactions scatter stellar mass compact objects onto a strong-field,
relativistic orbit of the black hole in a galaxy's center.  Further
evolution is then driven by GW emission.  If the black hole has a mass
of around $10^5 - 10^7\,M_\odot$, then these are targets for a
detector like eLISA.  The GWs that they generate can be heard out to
$z \sim 0.5 - 1$; we expect dozens to hundreds of events over a
space-based detector's mission lifetime\cite{emriupdate}.  Measuring
the GWs from these events will provide precision data on the
characteristics of the large black hole, on the small body's orbit,
and on the mass of the small body --- in short, a precision probe of
the astrophysical population of galactic center black holes, and
information about the population of stars in the centers of galaxies.
It is expected that EMRI waves will even be precise enough to test the
Kerr solution by mapping the multipolar structure of the dense, dark
objects in galactic cores that we presume to be general relativity's
black holes.

\section{Modeling EMRIs}

To achieve these science goals, we must accurately model EMRI waves.
How accurate must our models be?  The answer depends on the purpose to
which we put the model\cite{lob2008}.  The instantaneous EMRI wave
amplitude will typically be about a factor of 10 smaller than detector
noise.  By fitting to a model template that is coherent in phase with
the data for $N$ cycles, we (roughly speaking) boost the
signal-to-noise ratio (SNR) by $N^{1/2}$.  For {\it detection}
purposes (determining that a signal is in your data) your model should
hold phase with the signal to within $\Delta\phi \lesssim 1$ radian
over the signal's duration\footnote{Note that ``duration'' does not
  necessary mean the complete span of the signal in your data.  One
  can break the data into segments, coherently integrate each segment
  against a template, and combine each processed segment into a single
  statistic.}.  The best fit is likely to have large systematic
errors, but that is acceptable if our goal is just to establish that a
signal is present.  For {\it measurement} purposes (e.g., using the
detected wave to determine source parameters), our model must be
accurate enough that systematic errors (due to inadequate modeling)
are smaller than statistic errors (due to noise).  A crude rule of
thumb is that the template's phase must match the signal to within
$\Delta\phi \lesssim 1/\mbox{SNR}$.

Turn now to an overview of how one makes an EMRI model.  We will use
the action-angle approach described by Flanagan and
Hinderer\cite{hf2008} to describe the motion of the small body $m$ in
the spacetime of a larger black hole of mass $M$:
\begin{eqnarray}
\frac{dq_\alpha}{d\lambda} &=& \omega_\alpha({\bf J}) + \varepsilon
g_\alpha(q_\theta,q_r,{\bf J}) + O(\varepsilon^2)\;,
\label{eq:angleevolve}
\\
\frac{dJ_i}{d\lambda} &=& \varepsilon G_i(q_\theta, q_r, {\bf J})
+ O(\varepsilon^2)\;.
\label{eq:actionevolve}
\end{eqnarray}
In these equations, $\lambda$ is a time variable that is well adapted
to strong-field Kerr orbits, and $\varepsilon = m/M$.  The angle
variables
\begin{equation}
q_\alpha \doteq (q_t,q_r,q_\theta,q_\phi)
\label{eq:angles}
\end{equation}
each describe the motion of the small body about the black hole in
suitable coordinates; the action variables
\begin{equation}
J_i \doteq (E/m, L_z/m, Q/m^2)
\label{eq:actions}
\end{equation}
correspond to integrals of the motion that are conserved along a
``background'' orbit (i.e., in the limit of purely geodesic motion).
We will examine the forcing terms $g_\alpha$ and $G_i$ in more detail
momentarily.

To understand the small body's motion in this framework, let us
examine Eqs.\ (\ref{eq:angleevolve}) and (\ref{eq:actionevolve}) more
carefully.  At zeroth order in $\varepsilon$, these equations become
\begin{equation}
\frac{dq_\alpha}{d\lambda} = \omega_\alpha({\bf J})\;,\qquad
\frac{dJ_i}{d\lambda} = 0\;.
\label{eq:actionangleevolve0}
\end{equation}
In other words, when the $O(\varepsilon)$ corrections to the equations
of motion are not included, the angle variables accumulate at a rate
set by their associated frequency, and the integrals of the motion are
constant.  Equation (\ref{eq:actionangleevolve0}) expresses the fact
that the motion at zeroth order in the small body's mass is a Kerr
geodesic.

When we go to the next order in $\varepsilon$, the forcing terms
$g_\alpha$ and $G_i$ must be included:
\begin{equation}
\frac{dq_\alpha}{d\lambda} = \omega_\alpha({\bf J}) + \varepsilon
g_\alpha(q_\theta, q_r, {\bf J})\;,\qquad
\frac{dJ_i}{d\lambda} =
\varepsilon G_i(q_\theta, q_r, {\bf J})\;.
\label{eq:actionangleevolve1}
\end{equation}
These terms push the small body away from the geodesic, and constitute
the leading self force correction to the small body's motion.

\section{The two-timescale expansion}

Further insight into EMRI evolution can be found by separating the
forcing terms into their averages and oscillations about the average:
\begin{equation}
G_i(q_\theta, q_r, {\bf J}) = \langle G_i({\bf J})\rangle + \delta
G_i(q_\theta, q_r, {\bf J})\;,
\end{equation}
where
\begin{eqnarray}
\langle G_i({\bf J}) \rangle &=&
\frac{1}{(2\pi)^2}\int_0^{2\pi}dq_\theta \int_0^{2\pi}dq_\theta\,
G_i(q_\theta, q_r, {\bf J})\;,
\label{eq:avedef}
\\
\delta G_i(q_\theta, q_r, {\bf J}) &=& G_i(q_\theta, q_r, {\bf J}) -
\langle G_i({\bf J})\rangle\;.
\label{eq:oscdef}
\end{eqnarray}
We apply a similar split to $g_\alpha(q_\theta, q_r, {\bf J})$.
Rewrite the equations of motion once more:
\begin{eqnarray}
\frac{dq_\alpha}{d\lambda} &=& \omega_\alpha({\bf J}) + \varepsilon
\langle g_\alpha({\bf J})\rangle +
\varepsilon \delta g_\alpha(q_\theta, q_r, {\bf J})\;,
\label{eq:angleevolve2}
\\
\frac{dJ_i}{d\lambda} &=&
\varepsilon\langle G_i({\bf J})\rangle +
\varepsilon\delta G_i(q_\theta, q_r, {\bf J})\;.
\label{eq:actionevolve2}
\end{eqnarray}
The averaged forcing term $\langle G_i({\bf J})\rangle$ describes the
leading evolution of the small body's integrals of motion; its
components describe the dissipative evolution of $E$, $L_z$, and $Q$.
This term drives the secular evolution of the system's orbital
parameters on an inspiral time scale $T_{\rm i} \sim M/\varepsilon =
M^2/m$.  The averaged forcing term $\langle g_\alpha({\bf J})\rangle$
is equivalent to a shift of the frequencies:
\begin{equation}
\omega_\alpha({\bf J}) \longrightarrow \omega_\alpha({\bf J}) +
\varepsilon\langle g_\alpha({\bf J})\rangle.
\end{equation}
This shift is the leading conservative contribution of the small
body's self force.  These forcing terms are nearly constant, varying
on the long timescale $T_i \sim M^2/m$ that characterizes the rate of
change of the integrals of motion ${\bf J}$.  By contrast, the forcing
terms $\delta G_i(q_\theta, q_r, {\bf J})$ and $\delta
g_\alpha(q_\theta, q_r, {\bf J})$ vary rapidly on a timescale $T_{\rm
  o} \sim M$ that characterizes the small body's orbital motion.
Their impact is (usually) much less important than the impact of the
averaged terms $\langle G_i({\bf J})$ and $\langle g_\alpha({\bf
  J})\rangle$.

As discussed above, the most important detail we need to understand to
characterize models for GW measurements is the phase\footnote{Bear in
  mind that the following equation is meant to be schematic.  The
  frequency $\omega(t)$ which appears under this integral is in fact a
  harmonic of the various frequencies which characterize Kerr black
  hole orbits, and as such has more contributions than are indicated
  in this sketch.} accumulated over some interval:
\begin{equation}
\Phi(t_1,t_2) = \int_{t_1}^{t_2} \omega(t)\,dt = \Phi_{\rm diss-1} +
\Phi_{\rm cons-1} + \Phi_{\rm diss-2} + \Phi_{\rm cons-2} + \ldots
\end{equation}
The contributions to this phase have the following scalings with
masses, and arise from the following pieces of source physics:

\begin{itemize}

\item $\Phi_{\rm diss-1} = O(T_{\rm i}/T_{\rm o}) = O(M/m)$: The
  slowing evolving geodesic frequency $\omega = 1/T_{\rm o} \sim 1/M$
  integrated over the inspiral time $T_{\rm i} \sim M^2/m$.

\item $\Phi_{\rm cons-1} = O(\epsilon T_{\rm i}/T_{\rm o}) = O(1)$:
  The conservative correction to the frequency $\delta\omega \sim
  \varepsilon\omega$ integrated over the inspiral time $T_{\rm i}$.

\item $\Phi_{\rm diss-2} = O(\epsilon T_{\rm i}/T_{\rm o}) = O(1)$:
  The slowly evolving geodesic frequency $\omega$ integrated against
  the oscillatory correction to the inspiral time $\delta T_{\rm i}
  \sim \varepsilon T_{\rm i}$.

\item $\Phi_{\rm cons-2} = O(\epsilon^2 T_{\rm i}/T_{\rm o}) =
  O(m/M)$: The conservative correction to the frequency $\delta\omega
  \sim \varepsilon\omega$ integrated against the oscillatory
  correction to the inspiral time $\delta T_{\rm i} \sim \varepsilon
  T_{\rm i}$.

\end{itemize}
These schematic countings suggest that we must include the leading
adiabatic conservative piece and perhaps the first oscillatory
dissipative piece in order to have effective {\it detection}
templates.  We certainly will need to go farther for measurement
purposes, or else just accept that a certain level of systematic error
may be very difficult to remove from EMRI waveform models.

\section{Adiabaticity and its limitations: Resonant orbits}

\begin{figure}
\begin{center}
\includegraphics[width=2.4in]{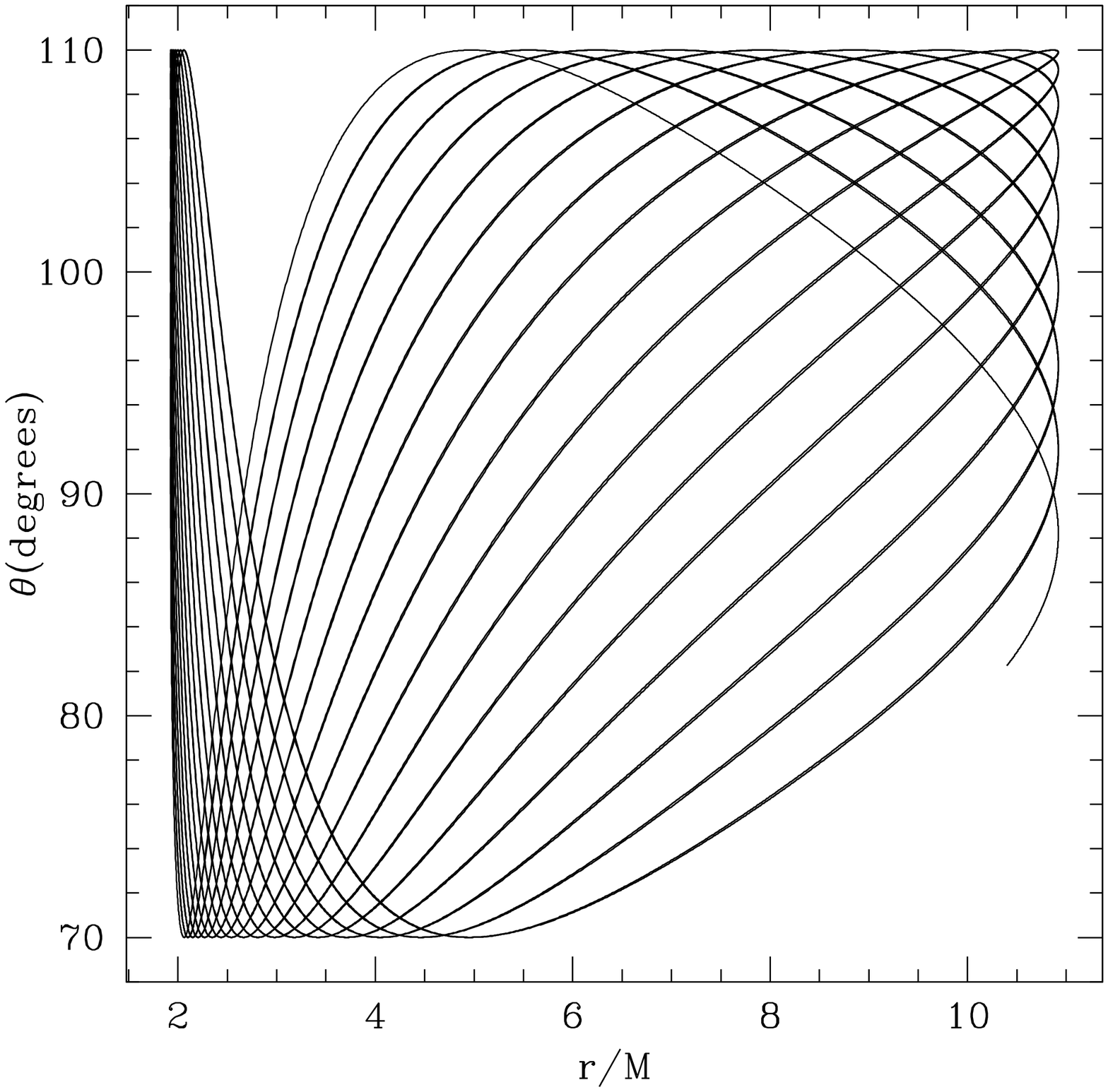}
\includegraphics[width=2.4in]{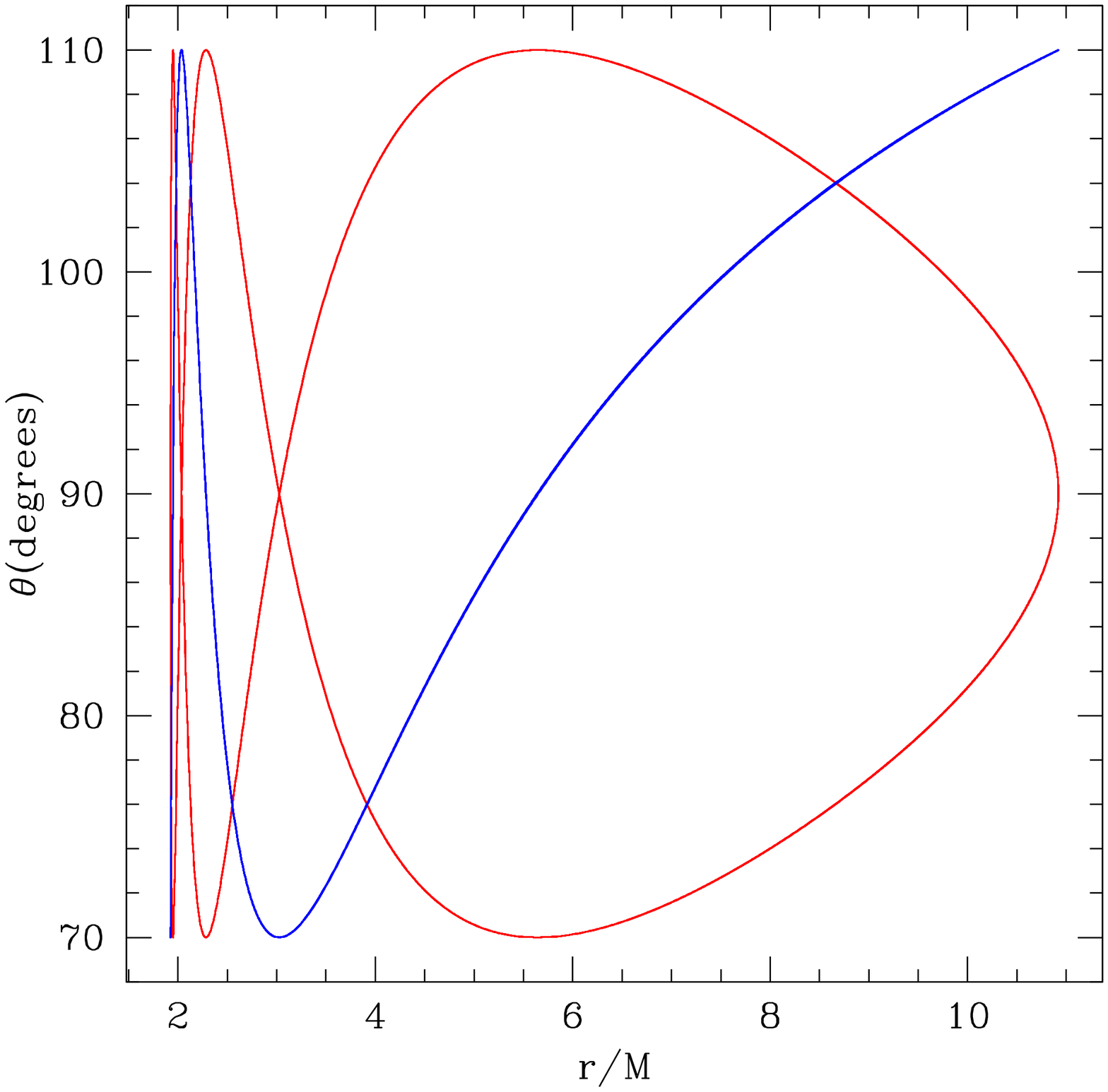}
\end{center}
\caption{Three example Kerr black hole orbits.  All oscillate between
  $2M \lesssim r \lesssim 12M$ and $70^\circ \le \theta \le
  110^\circ$; roughly 9 radial cycles are shown for each case.  The
  left-hand panel shows these orbits for a black hole with spin
  parameter $a = 0.95M$.  This motion is ergodic --- given enough
  time, the orbit would come arbirarily close to every accessible
  point in $(r,\theta)$.  The right-hand panel shows two orbits for
  spin $a = 0.9M$.  The $\theta$ and $r$ motions of these orbits are
  in a 3:1 resonance: the orbit oscillates three times in $\theta$
  over each radial cycle.  No matter how long we wait, the motion will
  be confined to the Lissajous figures shown here.  The particular
  points the orbit passes through depend on the system's phase as it
  enters resonance.  Two examples are shown here.}
\label{fig:ergo_liss}
\end{figure}

The fact that the oscillatory contributions to the EMRI model are
subleading suggests that a useful approximation may be to ignore them
at first pass.  Doing so gives us the {\it adiabatic approximation} to
EMRI evolution:
\begin{equation}
\frac{dq_\alpha}{d\lambda} = \omega_\alpha({\bf J}) + \varepsilon
\langle g_\alpha({\bf J})\rangle\;,\qquad
\frac{dJ_i}{d\lambda} =
\varepsilon\langle G_i({\bf J})\rangle\;.
\label{eq:adiabatic}
\end{equation}
For {\it most} black hole orbits, this approximation works well.  This
is because the small body's motion typically is ergodic: after a small
number of orbits (requiring far less than the inspiral time), the
small body has come close to every point it is allowed to pass
through.  Its motion thus averages the forcing terms in the equations
of motion more or less automatically.  This is illustrated in the
left-hand panel of Fig.\ {\ref{fig:ergo_liss}} (adapted from Fig.\ 1
of Ref.\ \refcite{fhr2014}), which shows the path in $(r,\theta)$
traced out by an orbit.  Roughly nine radial orbits are shown here.
Given enough time, this trace would fill the entire $(r,\theta)$ plane
over the domain $2M \lesssim r \lesssim 12M$, $70^\circ \le \theta \le
110^\circ$.

However, there exist orbits for which this averaging does not occur.
If the small body's $\theta$ and $r$ frequencies are in a low-order
resonance, then the motion is not ergodic, but instead traces out a
Lissajous figure in the $(r,\theta)$ plane.  An example is shown in
the right-hand panel of Fig.\ {\ref{fig:ergo_liss}}.  Because these
orbits do not come close to all allowed points in the accessible
physical space, they do not effectively average the forcing functions
$g_\alpha(q_\theta, q_r, {\bf J})$ and $G_i(q_\theta, q_r, {\bf J})$.
Indeed, different initial conditions trace out different Lissajous
figures.  This means that the detailed manner in which averaging fails
depends on an orbit's phase as it enters resonance.

To make this more quantitative, examine the Fourier expansion of
$G_i$:
\begin{equation}
G_i(q_\theta, q_r, {\bf J}) = \sum_{kn} G_{i;kn}({\bf J})
e^{i(kq_\theta + nq_r)}\;.
\label{eq:Fourierforcing}
\end{equation}
For a non-resonant orbit, only one term on the right-hand survives
averaging: using Eq.\ (\ref{eq:avedef}), we have
\begin{equation}
\langle G_i({\bf J})\rangle_{\rm non-res} = G_{i;00}({\bf J})\;.
\end{equation}
since $\langle e^{i(kq_\theta + nq_r)}\rangle = 0$ except for $k = n
= 0$.

This is not the case for resonant orbits.  When
$\Omega_\theta/\Omega_r$ are in a ratio of small integers, then
$q_\theta/q_r$ are also in a ratio of small integers.  We now find
that many terms on the right-hand side of
Eq.\ (\ref{eq:Fourierforcing}) survive the average:
\begin{equation}
\langle G_i({\bf J})\rangle_{\rm res} = G_{i;00}({\bf J})
+ \sum_{(k,n)} G_{i;kn}\;.
\end{equation}
The final sum is over all pairs $(k,n)$ that satisfy $kq_\theta + nq_r
= 0$.

Averaging, and thus the adiabatic approximation, fails as we enter a
resonance.  The oscillatory terms in the equations of motion combine
in phase on a resonance, ``kicking'' the system until it evolves away
from the resonance.  The adiabatic evolution does well at describing
the system's evolution before and after the resonance, but does a poor
job at modeling in the system very near the resonance.  Taking
resonances into account, we find that the system's phase evolution
must be modified:
\begin{eqnarray}
\Phi(t_1,t_2) &=& \int_{t_1}^{t_2} \omega(t)\,dt
\nonumber\\
&=& \Phi_{\rm diss-1} + \Phi_{\rm RES} +
\Phi_{\rm cons-1} +
\Phi_{\rm diss-2} +
\Phi_{\rm cons-2} + \ldots\;.
\end{eqnarray}
This form is identical to that given earlier, but there is now a new
term: $\Phi_{\rm RES} = O(\epsilon^{1/2} T_{\rm i}/T_{\rm o}) =
O([M/m]^{1/2})$.  {\it This term dominates over all but the leading
  dissipative contributions to the system's phase evolution.}
Detailed analysis\cite{fh2012,rh2014} indeed shows that $\Phi_{\rm
  RES}$ contributes dozens to hundreds of radians to the system's
phase evolution, substantially more than all terms except the leading
one.

One might imagine that, since resonant orbits are a set of measure
zero in the complete set of Kerr black hole orbits, these cases are
curiosities that are unlikely to play much role in astrophysics.  That
is not the case.  Consider a set of astrophysical inspirals with
parameters such that they are likely to be important sources for
low-frequency GW detector.  We have shown\cite{rh2014} that {\it
  every} inspiral will pass through at least one dynamically
significant low-order resonance as it spirals through the detector's
sensitive band.  Many of these inspirals will pass through two
significant resonances; some will pass through three.

\section{Summary and outlook}

Although useful for producing a somewhat accurate picture of extreme
mass ratio binaries, the adiabatic approach to inspiral is ultimately
inadequate for modeling these sources, even for the less stringent
task of developing detection templates.  We must go beyond this
picture and develop post-adiabatic EMRI models in order to more
completely model these sources.

Of particular importance is understanding the magnitude of the
``kick'' that is imparted to an EMRI's evolution by each resonance
passage.  Properly doing this requires that we self consistently
integrate the equations of motion, including the oscillatory part of
the self interaction.  Past work\cite{fh2012,fhr2014,rh2014} has given
us some idea how much kick we can expect and the number of cycles over
which the kick operates, but it remains important to develop a fully
self consistent inspiral and waveform model to assess the reliability
of these estimates.

Even with good modeling, it is likely that the impact of resonances
will substantially complicate our ability to measure EMRI GWs.  The
detailed evolution of an EMRI on resonance depends on the value of two
orbital phases as we enter resonance.  For non-resonant orbits, these
phases are ignorable.  Resonances thus increase the dimensionality of
the EMRI waveform manifold, and potentially greatly expand the number
of parameters that will be needed for measurement templates.

For detection purposes at least, it may be adequate to break the data
into segments.  A simple, adiabatic model suffices to model the
system's phase until we are within a few dozen or hundred radians of
the resonance; a different simple, adiabatic model suffices to model
the phase once we are a few dozen or hundred radians past resonance.
Each EMRI is thus broken into pre- and post-resonance segments.
Similar techniques are used in radio astronomy to model glitching
pulsars, when a pulsar's spin frequency suddenly changes following a
change in a neutron star's moment of inertia due to a rearrangement of
its internal fluid distribution (see, e.g., Fig.\ 1 of
Ref.\ {\refcite{link92}} for an example and discussion).  This
segmented approach is likely to work well on EMRI events whose fully
coherent signal-to-noise ratios are several tens or larger.

Much work remains to be done on this challenging problem.

\section*{Acknowledgments}

Research in our group is supported by National Science Grant
PHY-1403261.  The work discussed here has been done in collaboration
with \'Eanna Flanagan and Uchupol Ruangsri, and has benefitted greatly
from discussions with Tanja Hinderer, Eric Poisson, and Niels
Warburton.


\begin{thebibliography}{0}

\bibitem{numrel1} L.\ Lehner and F.\ Pretorius, Ann.\ Rev.\ Astron.
  Astrophys.\ {\bf 52}, 661 (2014).

\bibitem{numrel2} U.\ Sperhake, Class.\ Quantum Grav.\ {\bf 32},
  124011 (2015).

\bibitem{eob} T.\ Damour, in Procedings of the conference ``Relativity
  and Gravitational --- 100 years after Einstein in Prague,''
  arXiv:1212.3159.

\bibitem{eLISA} The eLISA Consortium (P.\ Amaro Seoane et al.), {\it
  The Gravitational Universe}, arXiv:1305.5720.

\bibitem{decigo} K.\ Yagi, Int.\ J.\ Mod.\ Phys.\ D {\bf 22}, 1341013
  (2013).

\bibitem{emriupdate} P.\ Amaro-Seoane, J.\ R.\ Gair, A.\ Pound,
  S.\ A.\ Hughes, and C.\ F.\ Sopuerta, J.\ of Phys.: Conf.\ Ser. {\bf
    610}, 012002 (2015).

\bibitem{lob2008} L.\ Lindblom, B.\ J.\ Owen, and D.\ A.\ Brown,
  Phys.\ Rev.\ D {\bf 78}, 124020 (2008).

\bibitem{hf2008} T.\ Hinderer and E.\ E.\ Flanagan, Phys.\ Rev.\ D
  {\bf 78} 064028 (2008).

\bibitem{fhr2014} E.\ E.\ Flanagan, S.\ A.\ Hughes, and U.\ Ruangsri,
  Phys.\ Rev.\ D {\bf 89}, 084028 (2014).

\bibitem{fh2012} E.\ E.\ Flanagan and T.\ Hinderer,
  Phys.\ Rev.\ Lett.\ {\bf 109}, 071102 (2012).

\bibitem{rh2014} U.\ Ruangsri and S.\ A.\ Hughes, Phys.\ Rev.\ D {\bf
  89}, 084036 (2014).

\bibitem{link92} B.\ Link, R.\ I.\ Epstein, and K.\ A.\ van Riper,
  Nature {\bf 359}, 616 (1992).

\end{thebibliography}
\end{document}